\documentclass[
aps, 
prb, 
10pt,  
amssymb, 
amsmath, 
superscriptaddress, 
tightenlines, 
twocolumn, 
notitlepage, 
] {revtex4-2}
\usepackage{color}
\usepackage{graphicx}
\usepackage{dcolumn}
\usepackage{bm}

\begin{document}



\title{Quantized nonlinear Hall effect from chiral monopole}

\author{Nikolai Peshcherenko}%
\affiliation{%
Max Planck Institute for Chemical Physics of Solids, 01187, Dresden, Germany}

\author{Claudia Felser}%
\affiliation{%
Max Planck Institute for Chemical Physics of Solids, 01187, Dresden, Germany}

\author{Yang Zhang}
\affiliation{Department of Physics and Astronomy, University of Tennessee, Knoxville, TN 37996, USA}
\affiliation{Min H. Kao Department of Electrical Engineering and Computer Science, University of Tennessee, Knoxville, Tennessee 37996, USA}



\begin{abstract}
Nonlinear Hall effect arises in materials without inversion symmetry, and the intrinsic contribution is typically from Berry curvature dipole of non-universal Fermi pockets. Here we propose that nonlinear Hall effect can reach quantization in chiral Weyl semimetals without mirror symmetries. The energy shift between a pair of Weyl nodes leads to chirally asymmetric intra-node relaxation, and the net trace of nonlinear Hall conductivity is thus quantized in units of $e^3/\hbar^2$ and determined by sum of monopole charge weighted by the transport relaxation time. Our theory also applies to mirror symmetric Weyl/Dirac semimetals with chiral anomaly. Additionally, besides DC transport probes, we anticipate that nonlinear circular dichroism measurements could detect chiral asymmetry-induced currents.
\end{abstract}

\maketitle

\section{Introduction} 
Over the past several decades, Hall effects have emerged as an ideal probe for studying topological phases of matter \cite{QH_topological1,QH_topological2,QH_topological3,QH_topological4,QH_topological5}. Most of known Hall effects result from broken time-reversal symmetry, induced either by magnetic fields \cite{Hall_magnetic}, magnetic orderings \cite{hall_magn1,hall_magn2,hall_magn3,hall_magn4,hall_magn5} or non-vanishing Berry curvature \cite{ahe_berry_curvature,ahe_berry_curvature2}. Nevertheless, according to Onsager relations \cite{onsager,TR_argument}, having finite Hall response is not prohibited in TR symmetric systems (with inversion symmetry broken nonetheless) if one goes beyond linear in electric field regime. Nonlinear Hall effects \cite{nonlinear_hall_effects} have thus been predicted theoretically \cite{Berry_curvature_dipole,zhang2018berry,nonlinear_hall,nonlinear_Hall1,nonlinear_Hall2,nonlinearHall_3,nonlinear_Hall4} and observed experimentally \cite{nonlinear_Hall6,nonlinear_Hall7}.

All Hall contributions could be conveniently separated \cite{hall_magn2} into intrinsic (i.e., determined by band topology) and extrinsic (given solely by impurity scattering) parts. The intrinsic part for TR symmetric system was previously demonstrated to originate predominantly from the so-called Berry curvature dipole contribution \cite{Berry_curvature_dipole}. This contribution is the strongest when linear crossing of conduction and valence bands occurs in three-dimensional (3D) momentum space close to Fermi level.  The crossings are referred to as Weyl points and are known \cite{berry_review} to act as monopoles of the Berry curvature, so that Berry curvature flux around Weyl nodes is quantized. Thus, one may expect a quantized Berry curvature-related response of electronic subsystem.

However, observing truly quantized features remains a challenging task. Although quantization of photogalvanic response was previously suggested \cite{de2017quantized}, the quantization was later proven to be non-robust with respect to electron-electron interactions \cite{Kozii,interactions}. Non-quantized is also Berry curvature dipole contribution mentioned above due to its dipole and not monopole nature. Thus, further studies in this domain may be in order. 

In this work, we propose that net nonlinear Hall current can reach quantization in the presence of chiral symmetry breaking. Crucial ingredients of our approach are the combination of anomalous velocity and shifted Fermi surface due to finite electric field. Since the Fermi surface shift itself is proportional to intra node momentum relaxation time, we prove that the result for non-linear Hall conductivity in Weyl semimetals is given by sum of Berry curvature flux-proportional terms, weighted by momentum relaxation times within each Weyl node. Thus, our setup for nonlinear Hall current observation requires an explicit chiral symmetry breaking provided by intrinsic mirror symmetry breaking \cite{de2017quantized,rees2020helicity,ni2021giant}. 

In extreme chiral limit, where remote Weyl nodes have vanishing relaxation times, nonlinear Hall current arises solely from Weyl monopoles near Fermi level, as a simplification of Berry curvature dipole mechanism. Furthermore, nonlinear Hall effect from dipole moment of Berry curvature \cite{Berry_curvature_dipole} requires a certain degree of electron's spectrum anisotropy (e.g., tilted Weyl nodes \cite{zhang2018berry}), while monopole nonlinear Hall effect from chiral asymmetry allows one to observe a topological monopole contribution not limited by extra anisotropy requirement. We further demonstrate that chiral asymmetry induced nonlinear Hall current could be also observed via nonlinear circular dichroism measurements. Calculations are made both within semiclassical Boltzmann equation and fully quantum approach using Kubo formula. Moreover, all of the physics discussed above applies to Dirac semimetals.

\section{Nonlinear Hall response}
\subsection{Semiclassical description}

\begin{figure}
    \centering
    \includegraphics[width=0.5\textwidth]{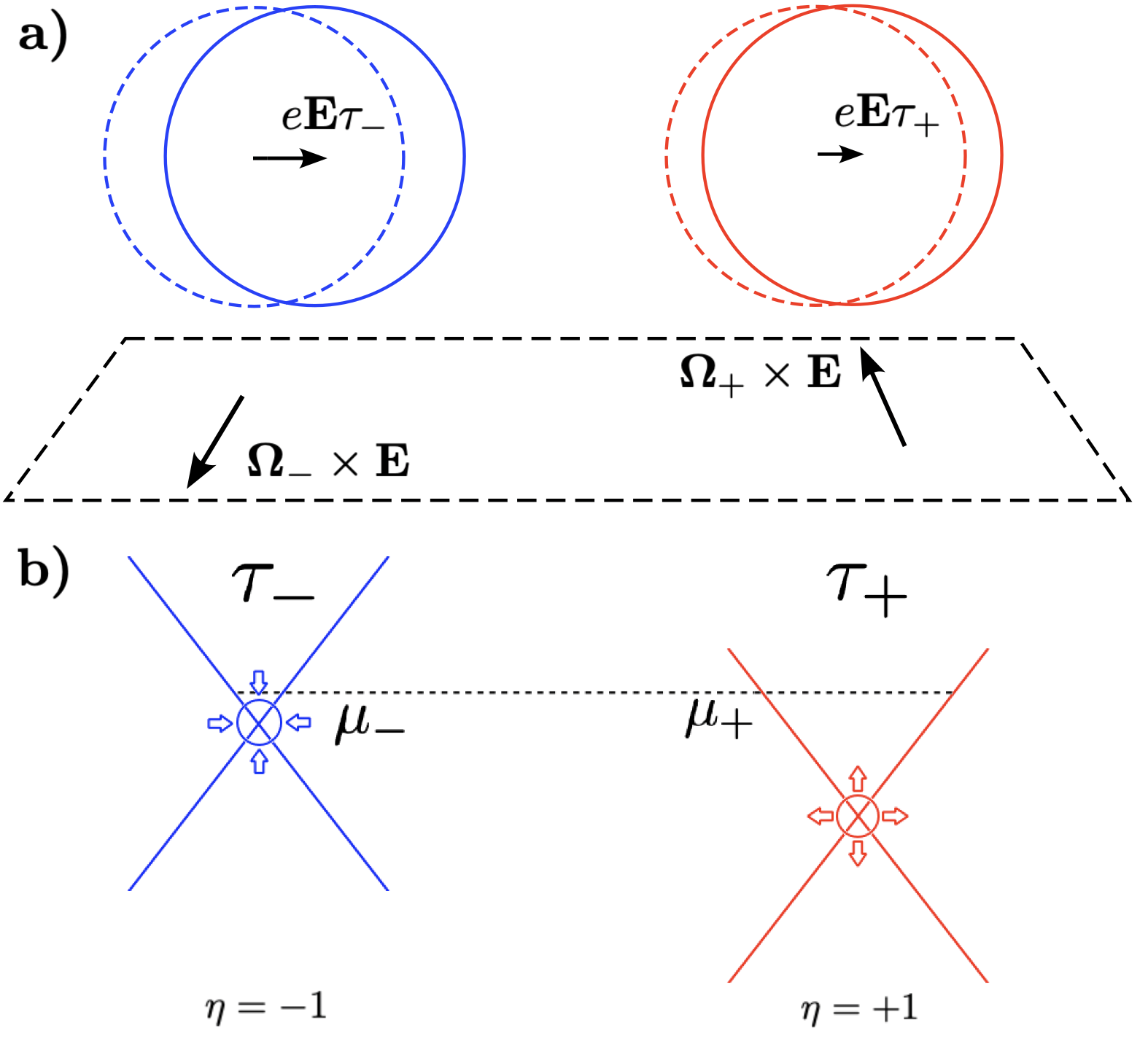}
    \caption{Schematic illustration of the nonlinear Hall effect origin. a) Fermi surface shift is shown to be proportional to node-dependent momentum relaxation time $\tau_+$, $\tau_-$. Anomalous Berry curvature velocities are shown to have opposite directions for opposite chirality nodes. b) 'Hot' and 'cold' Berry curvature monopoles that arise due to mirror symmetry breaking are shown. Chern number $\pm C$ is shown for each node ($C=4$ for CoSi).}
    \label{fig:pockets}
\end{figure}

General semiclassical expression for charge current $\mathbf{j}$ in topological Weyl semimetals is given by
\begin{align}
    \mathbf{j}=e\sum_\eta\sum_\mathbf{k}\mathbf{v}_\eta f_\eta(\mathbf{k}),\quad\mathbf{v}_\eta=\nabla_\mathbf{k}\varepsilon_\eta(\mathbf{k})+\mathbf{\Omega}_\eta(\mathbf{k})\times\dot{\mathbf{k}},\nonumber\\
    \dot{\mathbf{k}}=e\mathbf{E},
    \label{eq:semiclassical}
\end{align}
where electron velocity $\mathbf{v}_\eta$ of a band $\eta$ is given by a sum of group velocity $\nabla_\mathbf{k}\varepsilon_\eta(\mathbf{k})$ and anomalous velocity contribution $\mathbf{\Omega}_\eta\times\dot{\mathbf{k}}$. Electronic distribution function is described by $f_\eta(\mathbf{k})$. For calculation of distribution function $f_\eta=f_\mathrm{eq}+\delta f_\eta$ one could employ Boltzmann equation
\begin{align}
    -i\omega f_\eta+e\mathbf{E}(\omega)\cdot\nabla_\mathbf{k}f_\eta=-\frac{\delta f_\eta}{\tau_\eta},
\end{align}
so that its linear in $\mathbf{E}$ solution gives
\begin{align}
    \delta f_\eta(\omega_0)=-\frac{e\mathbf{E}_0\nabla_\mathbf{k}f_\mathrm{eq}}{-i\omega_0+1/\tau_\eta}
    \label{eq:delta_f}
\end{align}
where $\omega_0$ stands for frequency of a monochromatic light wave $\mathbf{E}(t)=\mathrm{Re}\{\mathbf{E}_0e^{-i\omega_0 t}\}$. Therefore, dc component of Hall current $j_H(\omega_j=0)$ is finally given by 
\begin{align}
    j_H(\omega_j=0)=-\frac{e^2}{\hbar^2}\sum_\eta\sum_\mathbf{k}\mathbf{\Omega}_\eta(\mathbf{k})\times \mathbf{E}^*_0\frac{e\mathbf{E}_0\nabla_\mathbf{k}f_\mathrm{eq}}{-i\omega_0+1/\tau_\eta}.
    \label{eq:Hcurr}
\end{align}
Note that in Eq. \eqref{eq:Hcurr} we deliberately dismissed linear in $\mathbf{E}$ contribution due to its proportionality to net Berry curvature $\sum_\eta\mathbf{\Omega}_\eta$, which is known to vanish in TR invariant systems. The second order response in Eq. \eqref{eq:Hcurr} accounts both for anomalous velocity and for shift of Fermi surface in the course of applied electric field $\mathbf{E}$ (see Fig. \ref{fig:pockets}a). In extreme driving limit where $\tau_{\eta}>>\tau_{-\eta}$, we have a net nonlinear dc current from Berry curvature monopole, as a reduction of so-called 'Berry curvature dipole' \cite{Berry_curvature_dipole} nonlinear Hall current. In our setup, it is crucial to have a node-dependent relaxation time $\tau_\eta$ as a probe of chirality imbalance.

Since for Weyl nodes surface integral of Berry curvature $\int\mathbf{\Omega}_\eta\cdot d\mathbf{S}_\mathbf{k}=2\pi\eta$ is quantized, for the trace of second order conductivity tensor $\sigma_{xyz}+\sigma_{yzx}+\sigma_{zxy}$ (we focus only on dc component of the response $j_x(\omega_j=0)=\sigma_{xyz}(0)E_{0y} E^*_{0z}$) we arrive at
\begin{align}
    \mathrm{Tr}\,\sigma_{\alpha\beta\gamma}(0)=\frac{e^3}{8\pi^2\hbar^2}\sum_\eta\frac{\eta\tau_\eta(\mu_\eta)}{1-i\omega_0\tau_\eta(\mu_\eta)},
    \label{eq:Weyl_results}
\end{align}
which is the main result of present work. Eq. \eqref{eq:Weyl_results} allows for a particularly simple interpretation for the case of low frequencies $\omega_0\tau_\eta\ll1$. Namely, topological current is proportional to Berry curvature and node's chirality $\eta$. However, at the same time it accounts for the Fermi surface shift, which is proportional to the shortest timescale of the problem ($\tau_\eta$). In case of broken chiral symmetry, the overall contribution is non-vanishing and $\mathrm{Tr}\sigma_{\alpha\beta\gamma}(0)\propto\tau_+-\tau_-$. This could be understood with the help of time-resolved Hall signal (see Fig. \ref{fig:dichroism}a). In high frequency regime, Fermi surface shift becomes frequency and not $\tau_\eta$ dependent, thus effectively restoring chiral symmetry. Therefore, nonlinear Hall current in Eq. \eqref{eq:Weyl_results} vanishes for $\omega_0\tau_\eta\gg1$. 

For short-ranged impurity scattering it was proven previously that $\tau^{-1}_\eta\propto \nu(\mu_\eta)\propto\mu_\eta^2$ \cite{scat_time}. Therefore, in extreme limit $\mu_-\ll\mu_+$ one should have a truly Berry monopole contribution, provided that $\tau_-\gg\tau_+$. This is indeed the case for CoSi due to its Weyl nodes being separated by $0.2\,\mathrm{eV}$. Hence, assuming typical momentum relaxation time $\tau\sim 10^{-13}\,s$ and extreme limit $\tau_-\gg\tau_+$, we have $\sigma_{xyz}\sim7\cdot10^{-4}\frac{A}{V^2}$. It is also worth mentioning that the required mirror symmetry breaking could arise due to chiral anomaly effect \cite{nielsen1983adler} in parallel electric and magnetic fields. However, typical separation of nodes in energy space is likely to be smaller. For instance, for TaAs subject to $B=9\,\mathrm{T}$, $E=10^4\,\mathrm{V/m}$, $\tau\sim10^{-13}\,\mathrm{s}$ and chiral relaxation time $\tau_\mathrm{ch}\sim 10^{-11}\,\mathrm{s}$ one may estimate that due to chiral anomaly effect $\mu_+-\mu_-\sim 2\,\mathrm{meV}$, employing  analytical results of \cite{SpivakSon,Weyl_Boltzmann_eq}.

\subsection{Quantum Kubo formula approach}
The result for nonlinear Hall response of Eq. \eqref{eq:Weyl_results} could be confirmed with the help of fully quantum Kubo formula computation. As we show in what follows, it can be applied to both Dirac and Weyl semimetals unlike semiclassical calculation.
The most general form of second order current response in Fourier representation reads
\begin{align}
    j_\alpha(\omega_j)=\int\frac{d\omega}{2\pi}\sigma_{\alpha\beta\gamma}(\omega,\omega_j-\omega)E^{\beta}(\omega)E^{\gamma}(\omega_j-\omega)
    \label{eq:Fourier_current}
\end{align}
where $\mathbf{E}=\mathrm{Re}\left\{\mathbf{E}_0e^{-i\omega_0 t}\right\}$.

\begin{figure}
    \centering
    \includegraphics[width=0.4\textwidth]{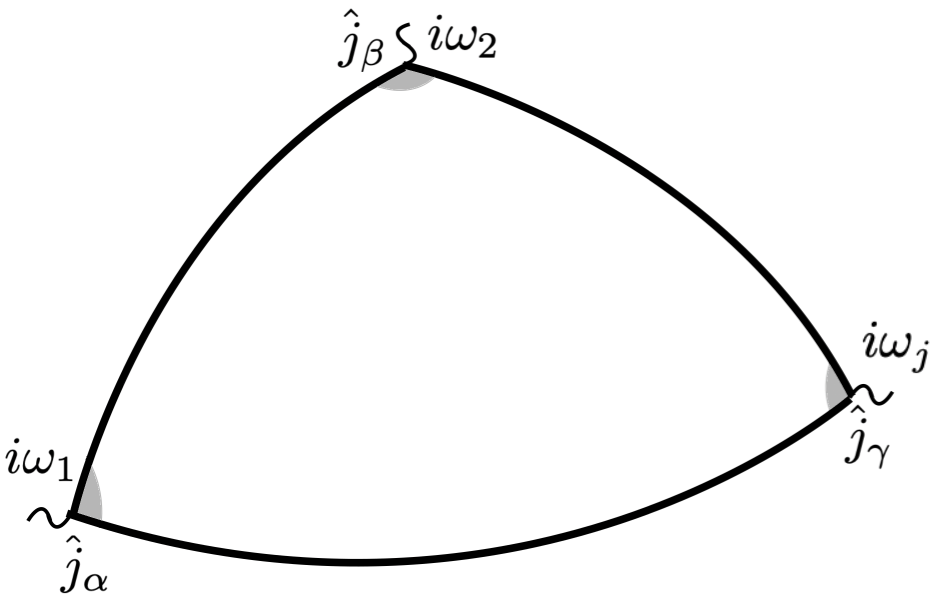}
    \caption{Feynman diagram for non-linear conductivity in Matsubara representation $\sigma_{\alpha\beta\gamma}(i\omega_1,i\omega_2)$, with $\omega_j=\omega_1+\omega_2$. Solid lines represent disorder-averaged Green function $\langle\hat{G}\rangle$ from Eq. \eqref{eq:Kubo}. Vertex corrections are shown by gray color.}
    \label{fig:diagram}
\end{figure}

In order to simplify second order response formula for $\sigma_{\alpha\beta\gamma}(\omega_1,\omega_2)$ we switch to Matsubara representation (see also \cite{Kozii,kozii2,Kubo_basis}and Fig. \ref{fig:diagram} for Feynman diagram representation):
\begin{align}
    \sigma_{\alpha\beta\gamma}(i\omega_1,i\omega_2)=\frac{\chi_{\alpha\beta\gamma}(i\omega_1,i\omega_2)+\chi_{\beta\alpha\gamma}(i\omega_2,i\omega_1)}{\omega_1\omega_2},\nonumber\\
    \chi_{\alpha\beta\gamma}(i\omega_1,i\omega_2)=\nonumber\\=\frac{1}{V}\int\frac{d\varepsilon}{2\pi}\mathrm{Tr}\left\langle\hat{j}_\alpha \hat{G}(i\varepsilon-i\omega_1)\hat{j}_\beta \hat{G}(i\varepsilon-i\omega_j)\hat{j}_\gamma \hat{G}(i\varepsilon)\right\rangle_\mathrm{dis},
    \label{eq:Kubo}
\end{align}
where $\omega_j=\omega_1+\omega_2$ and $\langle...\rangle_\mathrm{dis}$ is disorder average. For the sake of clarity we assume that electronic properties are governed by a standard spherically symmetric Weyl electron Hamiltonian 
\begin{align}
    \hat{H}=\eta v_F{\bm\sigma}\cdot\mathbf{k},
    \label{eq:Weyl_ham}
\end{align}
so that particle current $\hat{j}_\alpha$ and Green function $\hat{G}(i\varepsilon)$ are given by
\begin{align}
    \hat{j}_{\alpha}=e\partial_{k_\alpha}\hat{H}=\eta e\sigma_\alpha v_F,\quad \hat{G}(i\varepsilon)=(i\varepsilon-\hat{H})^{-1}.
\end{align}
Explicit computation of the non-linear response Eq.\eqref{eq:Kubo} (see supplemental material for details) leads to the same result of Eq. \eqref{eq:Weyl_results}. However, we would like to emphasize that Eq. \eqref{eq:Weyl_results} is justified only when interband optical transitions could be fully neglected (in the limit of $\omega_0\ll\mu_\eta$).

\section{Circular dichroism probe}
\begin{figure}
    \centering
    \includegraphics[width=0.5\textwidth]{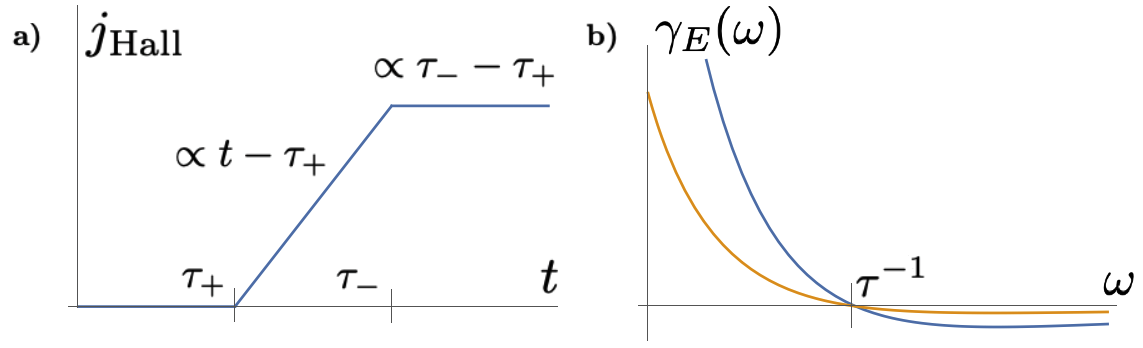}
    \caption{a) Time-resolved Hall current. $t<\tau_+$: ballistic currents in both $\eta=\pm C$ nodes compensate each other. $\tau_+<t<\tau_-$: $\eta=+C$ current is saturated while $\eta=-C$ is not. $\tau_-<t$: both $\eta=\pm1$ currents are saturated. b) Chiral dichroism signal $\gamma_E$ due to non-linear Hall response. Blue line: $(\tau_+-\tau_-)/\tau_0=0.2$, orange line: $(\tau_+-\tau_-)/\tau_0=0.15$, where $\tau_0=(\tau_++\tau_-)/2$.}
    \label{fig:dichroism}
\end{figure}

Non-trivial Hall response discussed above could also manifest itself in optical measurements of circular dichroism \cite{circular_dichroism1,circular_dichroism2,TR_breaking_CD}. For TR invariant systems with broken inversion symmetry, a homogeneous linear response is prohibited by Onsager symmetry relations \cite{TR_argument,onsager}. In previous works \cite{kargarian2015theory, dichroism}, consideration was mainly given to $\mathbf{q}$-linear contribution (with $\mathbf{q}$ being incident light wavevector). In contrast, here we would focus on non-linear part of the response instead. Employing the famous relation between conductivity $\sigma_{ij}(\omega_0)$ and dielectric $\varepsilon_{ij}(\omega_0)$ tensors
\begin{align}
    \varepsilon_{ij}(\omega_0,\mathbf{q})=\delta_{ij}+\frac{4\pi i\sigma_{ij}(\omega_0,\mathbf{q})}{\omega_0},
\end{align}
we have been able to calculate eigenvalues of longitudinal and transversal (with respect to the electric field direction $\mathbf{E}$) eigenmodes. These eigenvalues are, in turn, refraction indices and are given by $n^2_\parallel=\varepsilon_\parallel(\omega_0)$, $n^2_{L(R)}=\varepsilon_\perp(\omega_0)\pm4\pi\sigma_{xy}/\omega_0$, where we have assumed isotropic $\varepsilon_{ij}$ in $0xy$ plane for the sake of simplicity. Chiral dichroism signal $\gamma=\mathrm{Im}(n_L-n_R)$ could then be demonstrated to be (also plotted in Fig. \ref{fig:dichroism}b for different magnetic field values)
\begin{align}
    \gamma_E\approx\frac{e^3}{6\pi\omega_0\hbar^2}E\sum_\eta\frac{\eta\tau_\eta}{1+\omega_0^2\tau_\eta^2}\mathrm{Re}\frac{1}{\sqrt{\varepsilon_\perp(\omega_0)}}.
    \label{eq:circ_dichroism}
\end{align}

The complete derivation of \eqref{eq:circ_dichroism} could be found in supplemental material. Now let us compare the finding of Eq. \eqref{eq:circ_dichroism} with the previously known result for
$\mathbf{q}$-linear contribution to circular dichroism signal \cite{Weyl_CPGE_speculation}. Note that even though $\gamma_E$ and $\gamma_q$ pertain to different kind of response (dc vs ac response), it is still instructive to compare magnitudes of the signals for future experimental tests. According to Ref. \cite{Weyl_CPGE_speculation}, $\gamma_q$ reads
\begin{align}
    \gamma_q=\frac{4e^2}{3\pi\hbar^2c}\sum_\eta \eta\mu_\eta\tau_\eta\mathrm{Re}\frac{1}{\sqrt{\varepsilon_\perp(\omega_0)}},
\end{align}
so that $\gamma_E\sim \frac{eE c}{\mu\omega_0}\gamma_q$ and for real material estimate (CoSi) ($E=100\,\mathrm{V}/\mathrm{cm}$, $\omega_0=10\,\mathrm{THz}$) $\gamma_E/\gamma_q\sim10$. 
This in turn means that chiral asymmetry related dichroism signal is easier to observe in dc response setup. 

We now discuss possible Fermi arc surface states contribution that inevitably arises due to topological Berry curvatures. In the limit of relatively large crystal volume, it is natural to expect surface contribution to be vanishingly small. Besides that, Weyl materials with few nodes could host a surface that does not support arc states \cite{kargarian2015theory}. We will leave the detailed study on surface contributions to future work.

\section{Discussion}

For a complete discussion of the result in Eq. \eqref{eq:Weyl_results}, we compare it with previous relevant findings. Since the result Eq. \eqref{eq:Weyl_results} is proportional to Berry curvature related anomalous velocity, it is evidently intrinsic. Therefore, it would make more sense to perform a separate comparison of our findings with nonlinear intrinsic \cite{Berry_curvature_dipole,Burkov_Hall} and extrinsic (side jump and skew scattering) \cite{nonlinear_hall,quantum_nonlinear_hall} contributions.

\subsection{Comparison with intrinsic nonlinear Hall effects from other sources}
We start with other known intrinsic mechanisms. 
The most influential work on this topic is the so-called 'Berry curvature dipole' contribution \cite{Berry_curvature_dipole} already mentioned above. This effect is the strongest near charge neutrality point. However, a relevant limit for the present study is for Fermi level being high above in electron band, where the Berry curvature dipole decays rapidly and the two effects could thus be comparable. For instance, simulations performed for TaIrTe$_4$ compound \cite{kumar2021room} suggest that Berry curvature dipole could vanish for certain values of chemical potential. For Weyl semimetal with mirror symmetry, Berry curvature dipole induced nonlinear Hall tensor in  Eq. \eqref{eq:Weyl_results} is vanishing.

Another similar non-linear effect was predicted in \cite{Burkov_Hall} as a consequence of magnetic field-dependent part of Lorentz force combined with anomalous velocity contribution in tilted Weyl nodes. This effect appears to be proportional to the tilt velocity that is usually small. For instance, for real material TaAs \cite{exp_data1,exp_data2} an order of magnitude estimate for $E=100\,\mathrm{V}/\mathrm{cm}$, $B=9\,\mathrm{T}$ gives $j_H/j_D\sim 1\%$ in \cite{Burkov_Hall}, which is less than our prediction, which would make $j_H/j_D\lesssim 5\%$ for TaAs. Furthermore, both mechanisms discussed above \cite{Berry_curvature_dipole,Burkov_Hall} heavily rely on presence of finite tilting and vanish for untilted nodes. Our result is free from such an assumption, due to its monopole nature.

\subsection{Comparison with extrinsic contributions}
Hall current could be also contributed to by asymmetrical impurity scattering (skew scattering) or a side jump \cite{quantum_nonlinear_hall} process. Let us start with nonlinear skew scattering contribution. It was previously studied for a TR-invariant materials in \cite{skew_liang_fu} and was estimated to be
\begin{align}
    j_\mathrm{sk}\sim env_F\left(\frac{eE\tau}{p_F}\right)^2\frac{\tau}{\Tilde{\tau}},
    \label{eq:skew}
\end{align}
where $\tau$, $\Tilde{\tau}$ stand for symmetric and asymmetric part of scattering rate correspondingly. One can compare it to the Drude current $j_D$
\begin{align}
    j_D\sim env_F\frac{eE\tau}{p_F}
\end{align}
so that
\begin{align}
    \frac{j_\mathrm{sk}}{j_D}\propto\frac{eE\tau}{p_F}\frac{\tau}{\Tilde{\tau}}.
\end{align}
For typical parameter values in Weyl semimetals $\tau\sim 10^{-13}\,s$, $\tau/\Tilde{\tau}\sim0.1$, $E=10^4\,\mathrm{V}/\mathrm{m}$ and $v_F=3\cdot10^5\,\mathrm{m}/\mathrm{s}$ one can estimate that $j_\mathrm{sk}/j_D\sim0.3\%$, which is significantly smaller than the intrinsic contribution and could be neglected. Smallness of the result is guaranteed by the smallness of skew scattering amplitude. 

For side jump scattering, a previous model study has revealed \cite{quantum_nonlinear_hall} that the contribution itself could be of the same order with intrinsic. Besides that, both contributions (intrinsic monopole one and side jump) demonstrate the same dependence on momentum relaxation time $\tau$. This is also the case in linear regime \cite{hall_magn2} and is the main reason why separation of these two contributions remains a challenging task. Furthermore, side jump contribution could depend sensitively on details of disorder potential (e.g., its rotation symmetry), which so far remain unknown in real materials. We therefore leave more accurate discussion of nonlinear side-jump contribution for future studies.

\subsection{Electron interaction and current quantization} For interband transitions \cite{optical_transition,Vanishing_response,ab_initio_PGE}, de Juan et al. predicted that photocurrent is quantized as a function of frequency \cite{optical_transition}. However, later it was proven that the presence of interaction could remove ideal quantization \cite{Kozii,interactions}. Since the nonlinear Hall current in Eq. \eqref{eq:Weyl_results} is sensitive only to chirally asymmetric scattering channels, the proposed quantization is therefore more robust under electron interactions.

\subsection{Dirac semimetal case} It is important to mention that despite the vanishing Berry curvature, all of the above physics could be applied to Dirac semimetals subject to parallel electric and magnetic fields. As demonstrated in \cite{Dirac_Andreev}, physics of Dirac material conductivity in this case is very similar to chiral anomaly in Weyl semimetal. Namely, node chirality in Weyl case is replaced by helicity for Dirac metal, with helicity being simply a projection of particle's spin onto its momentum. Parallel electric and magnetic fields would lead to an imbalance between different chirality/helicity populations.
For a more formal Kubo formula proof for nonlinear case please see supplemental material.


\section{Conclusions}
To summarise, we predicted a non-vanishing Hall response in TR-invariant systems from Berry curvature monopole. We have demonstrated that in the presence of mirror symmetry breaking  Weyl cones are shifted in energy space, which gives rise to chirally asymmetric intranode momentum relaxation times. 
Due to this asymmetry, nonlinear Hall current is given by quantized monopole charge, weighted by node specific relaxation time. We further discuss the corresponding  nonlinear circular dichroism signal from chiral asymmetry. The predicted effect is robust to interactions beyond simple impurity scattering, since only scattering channels which are chirally asymmetric would contribute to this effect. Our results could be used as a probe for the extent of chiral symmetry breaking. 

\textit{Acknowledgement.}
We are grateful to Philip Moll and Liang Fu for helpful discussions. C.F. is supported by the Catalyst Fund of Canadian Institute for Advanced Research. Y. Z. is supported by the start-up fund at University of Tennessee Knoxville. 

\nocite{*}

\bibliography{sample}

\begin{widetext}
\appendix
\section{Non-linear Hall response $\sigma_{\alpha\beta\gamma}$: Kubo formula derivation \label{Kubo_computation}}
In this supplemental material we explicitly derive the non-linear Hall response $\sigma_{\alpha\beta\gamma}$ for both Weyl and Dirac semimetals with the help of fully quantum Kubo formula. As shown in previous studies \cite{Kubo_basis,Kozii}, the general second order response $\sigma_{\alpha\beta\gamma}(i\omega_1,i\omega_2)$ is expressed through 3-point correlation functions $\chi_{\alpha\beta\gamma}(i\omega_1,i\omega_2)$: 
\begin{align}
    \chi_{\alpha\beta\gamma}(i\omega_1,i\omega_2)=\int\frac{d\varepsilon}{2\pi}\mathrm{Tr}\left\langle j_\alpha G(i\varepsilon-i\omega_1)j_\beta G(i\varepsilon-i\omega_j) j_\gamma G(i\varepsilon)\right\rangle_\mathrm{dis}.
    \label{eq:chi_definition}
\end{align}
Since the scattering is treated perturbatively, introducing disorder would lead only to a substitute of bare Green function by an averaged one plus vertex corrections. Disorder-averaged Green function $\langle\hat{G}\rangle$ of Weyl fermion in momentum representation is given by
\begin{align}
    \langle G\rangle(i\varepsilon,\mathbf{k})=\frac{P_+(\mathbf{k})}{i\varepsilon-v_Fk+\mu_\eta+\frac{i}{2\tau^{\prime}_\eta}\mathrm{sign}\varepsilon}+\frac{P_-(\mathbf{k})}{i\varepsilon+v_Fk+\mu_\eta+\frac{i}{2\tau^{\prime}_\eta}\mathrm{sign}\varepsilon},\quad P_\pm(\mathbf{k})=\frac{1}{2}\left(1\pm{\bm \sigma}\cdot\hat{\mathbf{k}}\right),
    \label{eq:G_simplified}
\end{align}
where $\mu_\eta$, $\tau^{\prime}_\eta=\frac{2}{3}\tau_\eta$ are the chirality-dependent chemical potential and scattering time correspondingly. In what follows we demonstrate that simple scattering time $\tau^{\prime}_\eta$ in Eq. \eqref{eq:G_simplified} differs from momentum relaxation time $\tau_\eta$ used in semiclassical analysis of Eq. (2) only by numerical factor $2/3$. It should be however noted that this relaxation times relation was proven previously in \cite{Dirac_Andreev} for the case of linear response. We generalize this to non-linear case.

We proceed with explicitly evaluating 3-point correlation function with disorder averaged Green function with later showing that vertex corrections will only lead to replacing of scattering time $\tau^{\prime}_\eta$ by transport momentum relaxation time $\tau_\eta$. For analytical derivation present in this section we stick to the approach developed in \cite{Kozii,kozii2} but with an account for impurity scattering.


\subsection{3-point correlation function $\chi_{\alpha\beta\gamma}$ calculation}
Since $j_\alpha=\eta ev_F\sigma_\alpha$, making use of Eq. \eqref{eq:chi_definition} and Eq. \eqref{eq:G_simplified}, for $\chi_{\alpha\beta\gamma}(i\omega_1,i\omega_2)$ we arrive at
\begin{align}
    \chi_{\alpha\beta\gamma}(i\omega_1,i\omega_2)=\eta e^3v_F^3\int\frac{d\varepsilon}{2\pi}\int\frac{d^3\mathbf{k}}{(2\pi)^3}\nonumber\\\bigg[\frac{\mathrm{tr}\left\{\sigma_\alpha P_-(\mathbf{k})\sigma_\beta P_+(\mathbf{k}) \sigma_\gamma P_+(\mathbf{k}) \right\}}{\left(i\varepsilon-i\omega_1+\mu+v_Fk+\frac{i}{2\tau^{\prime}_\eta}\mathrm{sign}(\varepsilon-\omega_1)\right)\left(i\varepsilon-i\omega_j-v_Fk+\mu+\frac{i}{2\tau^{\prime}_\eta}\mathrm{sign}(\varepsilon-\omega_j)\right)\left(i\varepsilon-v_Fk+\mu+\frac{i}{2\tau^{\prime}_\eta}\mathrm{sign}\varepsilon\right)}+\nonumber\\+\frac{\mathrm{tr}\left\{\sigma_\alpha P_+(\mathbf{k})\sigma_\beta P_-(\mathbf{k}) \sigma_\gamma P_-(\mathbf{k}) \right\}}{\left(i\varepsilon-i\omega_1-v_Fk+\mu+\frac{i}{2\tau^{\prime}_\eta}\mathrm{sign}(\varepsilon-\omega_1)\right)\left(i\varepsilon-i\omega_j+v_Fk+\mu+\frac{i}{2\tau^{\prime}_\eta}\mathrm{sign}(\varepsilon-\omega_j)\right)\left(i\varepsilon+v_Fk+\mu+\frac{i}{2\tau^{\prime}_\eta}\mathrm{sign}\varepsilon\right)}+\nonumber\\+\frac{\mathrm{tr}\left\{\sigma_\alpha P_+(\mathbf{k})\sigma_\beta P_+(\mathbf{k}) \sigma_\gamma P_-(\mathbf{k}) \right\}}{\left(i\varepsilon-i\omega_1-v_Fk+\mu+\frac{i}{2\tau^{\prime}_\eta}\mathrm{sign}(\varepsilon-\omega_1)\right)\left(i\varepsilon-i\omega_j-v_Fk+\mu+\frac{i}{2\tau^{\prime}_\eta}\mathrm{sign}(\varepsilon-\omega_j)\right)\left(i\varepsilon+v_Fk+\mu+\frac{i}{2\tau^{\prime}_\eta}\mathrm{sign}\varepsilon\right)}+\nonumber\\+\frac{\mathrm{tr}\left\{\sigma_\alpha P_-(\mathbf{k})\sigma_\beta P_-(\mathbf{k}) \sigma_\gamma P_+(\mathbf{k}) \right\}}{\left(i\varepsilon-i\omega_1+v_Fk+\mu+\frac{i}{2\tau^{\prime}_\eta}\mathrm{sign}(\varepsilon-\omega_1)\right)\left(i\varepsilon-i\omega_j+v_Fk+\mu+\frac{i}{2\tau^{\prime}_\eta}\mathrm{sign}(\varepsilon-\omega_j)\right)\left(i\varepsilon-v_Fk+\mu+\frac{i}{2\tau^{\prime}_\eta}\mathrm{sign}\varepsilon\right)}+\nonumber\\+\frac{\mathrm{tr}\left\{\sigma_\alpha P_+(\mathbf{k})\sigma_\beta P_-(\mathbf{k}) \sigma_\gamma P_+(\mathbf{k}) \right\}}{\left(i\varepsilon-i\omega_1-v_Fk+\mu+\frac{i}{2\tau^{\prime}_\eta}\mathrm{sign}(\varepsilon-\omega_1)\right)\left(i\varepsilon-i\omega_j+v_Fk+\mu+\frac{i}{2\tau^{\prime}_\eta}\mathrm{sign}(\varepsilon-\omega_j)\right)\left(i\varepsilon-v_Fk+\mu+\frac{i}{2\tau^{\prime}_\eta}\mathrm{sign}\varepsilon\right)}+\nonumber\\+\frac{\mathrm{tr}\left\{\sigma_\alpha P_-(\mathbf{k})\sigma_\beta P_+(\mathbf{k}) \sigma_\gamma P_-(\mathbf{k}) \right\}}{\left(i\varepsilon-i\omega_1+v_Fk+\mu+\frac{i}{2\tau^{\prime}_\eta}\mathrm{sign}(\varepsilon-\omega_1)\right)\left(i\varepsilon-i\omega_j-v_Fk+\mu+\frac{i}{2\tau^{\prime}_\eta}\mathrm{sign}(\varepsilon-\omega_j)\right)\left(i\varepsilon+v_Fk+\mu+\frac{i}{2\tau^{\prime}_\eta}\mathrm{sign}\varepsilon\right)}\bigg],
    \label{eq:master_chi}
\end{align}
where $\omega_j=\omega_1+\omega_2$, the sum over Landau level index $n$ is replaced by integration over semiclassical momentum $\mathbf{k}$ and $\mathrm{tr}\left\{...\right\}$ stands for the trace over spin degrees of freedom. In what follows we switch to integration in spherical coordinates of $\mathbf{k}$: $\int\frac{d^3\mathbf{k}}{(2\pi)^3}=\int_0^{\infty}\nu(\varepsilon) d\varepsilon\int\frac{d\Omega_\mathbf{k}}{4\pi}$, $\varepsilon=v_F|\mathbf{k}|$. As for angle integration, a straightforward calculation suggests that
\begin{align}
    \langle\mathrm{tr}\left\{\sigma_\alpha P_+(\mathbf{k})\sigma_\beta P_+(\mathbf{k}) \sigma_\gamma P_-(\mathbf{k}) \right\}\rangle_{\hat{\mathbf{k}}}=\langle\mathrm{tr}\left\{\sigma_\alpha P_+(\mathbf{k})\sigma_\beta P_-(\mathbf{k}) \sigma_\gamma P_-(\mathbf{k}) \right\}\rangle_{\hat{\mathbf{k}}}=\frac{1}{3}i\varepsilon_{\alpha\beta\gamma},
    \label{eq:identity}
\end{align}
where $\langle...\rangle_{\hat{\mathbf{k}}}$ stands for angular average over momentum directions. Thus, according to Eq. \eqref{eq:identity}, all traces appearing in \eqref{eq:master_chi} are identical which allows to effectively extend the integration over $\varepsilon$ to the whole real axis by grouping 2 subsequent terms into one contribution.
Assuming the limit $\omega_{1,2}\ll\mu$ (which is equivalent to neglecting all vertical optical transitions), we then arrive at
\begin{align}
    \chi_{\alpha\beta\gamma}(i
    \omega_1, i\omega_2)=\nonumber\\=\frac{1}{3}\varepsilon_{\alpha\beta\gamma}\eta e^3v_F^3\nu_F\int d\varepsilon\Bigg[\frac{\theta(\varepsilon)-\theta(\varepsilon-\omega_j)}{\left(2i\varepsilon-i\omega_1+2\mu+\frac{i}{2\tau^{\prime}_\eta}(\mathrm{sign}(\varepsilon-\omega_1)+\mathrm{sign}\varepsilon)\right)\left(-i\omega_j+\frac{i}{2\tau^{\prime}_\eta}(\mathrm{sign}(\varepsilon-\omega_j)-\mathrm{sign}\varepsilon)\right)}+\nonumber\\+\frac{\theta(\varepsilon-\omega_1)-\theta(\varepsilon-\omega_j)}{\left(2i\varepsilon-i\omega_1+2\mu+\frac{i}{2\tau^{\prime}_\eta}(\mathrm{sign}\varepsilon+\mathrm{sign}(\varepsilon-\omega_1))\right)\left(-i\omega_j+i\omega_1+\frac{i}{2\tau^{\prime}_\eta}(\mathrm{sign}(\varepsilon-\omega_j)-\mathrm{sign}(\varepsilon-\omega_1))\right)}\nonumber\\+\frac{\theta(\varepsilon)-\theta(\varepsilon-\omega_1)}{\left(-i\omega_1+\frac{i}{2\tau^{\prime}_\eta}(\mathrm{sign}(\varepsilon-\omega_1)-\mathrm{sign}\varepsilon)\right)\left(2i\varepsilon-i\omega_j+2\mu+\frac{i}{2\tau^{\prime}_\eta}(\mathrm{sign}(\varepsilon-\omega_j)+\mathrm{sign}\varepsilon)\right)}\Bigg]
\end{align}
Assuming for the sake of simplicity that $\omega_1>0$ and performing integration over $\varepsilon$, we have
\begin{align}
    \chi_{\alpha\beta\gamma}(i\omega_1,i\omega_2)\approx\eta e^3v_F^3\nu_F\frac{1}{3}\varepsilon_{\alpha\beta\gamma}\bigg[\frac{\omega_j}{2\mu(-i\omega_j-i/\tau^{\prime}_\eta)}+\frac{\omega_2}{2\mu(-i\omega_2-i/\tau^{\prime}_\eta)}+\frac{\omega_1}{2\mu(-i\omega_1-i/\tau^{\prime}_\eta)}\bigg]
\end{align}
However, it is evident that in the leading approximation in $\omega_{1,2}/\mu$ the expression $\chi_{\alpha\beta\gamma}(i\omega_1,i\omega_2)-\chi_{\alpha\beta\gamma}(i\omega_2,i\omega_1)$ vanishes. Therefore, one should take into account correction terms, keeping in mind the limit $\omega_1+\omega_2\to0$ and expanding over $\omega_{1,2}/\mu$:
\begin{align}
    \chi_{\alpha\beta\gamma}(i\omega_1,i\omega_2)\approx\eta e^3v_F^3\nu_F\frac{1}{3}\varepsilon_{\alpha\beta\gamma}\bigg[\frac{\omega_j}{2\mu(-i\omega_j-i/\tau^{\prime}_\eta)}+\frac{\omega_2}{2\mu(-i\omega_2-i/\tau^{\prime}_\eta)}+\frac{\omega_1}{2\mu(-i\omega_1-i/\tau^{\prime}_\eta)}+\frac{\omega_1^2}{4\mu^2(\omega_1+1/\tau^{\prime}_\eta)}\bigg]
    \label{eq:chi_result}
\end{align}
For $\chi_{\alpha\beta\gamma}(i\omega_2,i\omega_1)$ we have
\begin{align}
    \chi^{\alpha\beta\gamma}(i\omega_2,i\omega_1)\approx\eta e^3v_F^3\nu_F\frac{1}{3}\varepsilon_{\alpha\beta\gamma}\bigg[\frac{\omega_j}{2\mu(-i\omega_j-i/\tau^{\prime}_\eta)}+\frac{\omega_2}{2\mu(-i\omega_2-i/\tau^{\prime}_\eta)}+\frac{\omega_1}{2\mu(-i\omega_1-i/\tau^{\prime}_\eta)}-\frac{\omega_1^2}{4\mu^2(\omega_1+1/\tau^{\prime}_\eta)}\bigg]
\end{align}
so that for the chiral one node contribution we arrive at
\begin{align}
    \sigma_\mathrm{\alpha\beta\gamma}=\varepsilon_{\alpha\beta\gamma}\frac{1}{\omega^2}\eta e^3v_F^3\nu_F\frac{1}{3}\frac{\omega^2}{2\mu^2(-i\omega+1/\tau^{\prime}_\eta)}=\varepsilon_{\alpha\beta\gamma}\frac{1}{\omega^2}\eta e^3v_F^3\frac{\mu^2}{2\pi^2v_F^3}\frac{1}{3}\frac{\omega^2}{2\mu^2(-i\omega+1/\tau^{\prime}_\eta)}\to\varepsilon_{\alpha\beta\gamma}\eta\frac{e^3}{12\pi^2}\frac{\tau^{\prime}_\eta}{1-i\omega\tau^{\prime}_\eta}.
\end{align}
Thus, net nonlinear response reads (in full agreement with semiclassical result Eq. (4))
\begin{align}
    \sigma_\mathrm{\alpha\beta\gamma}=\varepsilon_{\alpha\beta\gamma}\frac{e^3}{12\hbar^2\pi^2}\sum_\eta\frac{\eta\tau^{\prime}_\eta}{1-i\omega\tau^{\prime}_\eta},
\end{align}
where $\eta$ stands for chirality. Note that extra factor of 2 arises due to different (again, by factor of 2) definitions of $\sigma_{\alpha\beta\gamma}$ in Eqs. (4) and (12).

\subsection{Vertex corrections}
Taking into account vertex corrections leads to the replacement of scattering time by transport momentum relaxation time. Since for Gaussian disorder in Weyl semimetals  $\langle V(\mathbf{r})V(\mathbf{r'})\rangle=n_\mathrm{imp}V^2\delta(\mathbf{r}-\mathbf{r'})=\frac{1}{\pi\nu_F\tau^{\prime}_\eta}(\mathbf{r}-\mathbf{r'})$, one ladder step contribution is given by
\begin{align}
    \frac{1}{\pi\nu_F\tau^{\prime}_\eta}\int\frac{d^3\mathbf{q}}{(2\pi)^3}\hat{G}(i\varepsilon-i\omega,\mathbf{q})\hat{j}_\alpha\hat{G}(i\varepsilon,\mathbf{q})=\frac{1}{\pi\nu_F\tau^{\prime}_\eta}\int_0^{\infty}\nu(\xi_\mathbf{q}) d\xi_\mathbf{q}\int \frac{d\mathbf{n_q}}{4\pi}\hat{G}(i\varepsilon-i\omega,\mathbf{q})\hat{j}_\alpha\hat{G}(i\varepsilon,\mathbf{q}),
    \label{eq:ladder_step}
\end{align}   
where $\xi_\mathbf{q}=v_F|\mathbf{q}|$. Plugging Eq. \eqref{eq:G_simplified} into Eq. \eqref{eq:ladder_step}, we arrive at
\begin{align}    
    \frac{1}{\pi\nu_F\tau^{\prime}_\eta}\int_0^{\infty} \nu(\xi_\mathbf{q}) d\xi_\mathbf{q} \int \frac{d\mathbf{n_q}}{4\pi} \frac{P_+(\mathbf{n_q})\hat{j}_\alpha P_+(\mathbf{n_q})}{(i\varepsilon-i\omega-v_Fq+\mu_\eta+\frac{i}{2\tau^{\prime}_\eta}\mathrm{sign}(\varepsilon-\omega))(i\varepsilon-v_Fq+\mu_\eta+\frac{i}{2\tau^{\prime}_\eta}\mathrm{sign}\varepsilon)}+\nonumber\\+\frac{1}{\pi\nu_F\tau^{\prime}_\eta}\int_0^{\infty} \nu(\xi_\mathbf{q}) d\xi_\mathbf{q} \int \frac{d\mathbf{n_q}}{4\pi} \frac{P_-(\mathbf{n_q})\hat{j}_\alpha P_-(\mathbf{n_q})}{(i\varepsilon-i\omega+v_Fq+\mu_\eta+\frac{i}{2\tau^{\prime}_\eta}\mathrm{sign}(\varepsilon-\omega))(i\varepsilon+v_Fq+\mu_\eta+\frac{i}{2\tau^{\prime}_\eta}\mathrm{sign}\varepsilon)}+\nonumber\\+\frac{1}{\pi\nu_F\tau^{\prime}_\eta}\int_0^{\infty} \nu(\xi_\mathbf{q}) d\xi_\mathbf{q} \int \frac{d\mathbf{n_q}}{4\pi} \frac{P_+(\mathbf{n_q})\hat{j}_\alpha P_-(\mathbf{n_q})}{(i\varepsilon-i\omega-v_Fq+\mu_\eta+\frac{i}{2\tau^{\prime}_\eta}\mathrm{sign}(\varepsilon-\omega))(i\varepsilon+v_Fq+\mu_\eta+\frac{i}{2\tau^{\prime}_\eta}\mathrm{sign}\varepsilon)}+\nonumber\\+\frac{1}{\pi\nu_F\tau^{\prime}_\eta}\int_0^{\infty} \nu(\xi_\mathbf{q}) d\xi_\mathbf{q} \int \frac{d\mathbf{n_q}}{4\pi} \frac{P_-(\mathbf{n_q})\hat{j}_\alpha P_+(\mathbf{n_q})}{(i\varepsilon-i\omega+v_Fq+\mu_\eta+\frac{i}{2\tau^{\prime}_\eta}\mathrm{sign}(\varepsilon-\omega))(i\varepsilon-v_Fq+\mu_\eta+\frac{i}{2\tau^{\prime}_\eta}\mathrm{sign}\varepsilon)}
    \label{eq:ladder_calculation}
\end{align}
Note however that the last two terms in Eq. \eqref{eq:ladder_calculation} would be suppressed by small parameter $\omega/\mu\ll1$ and could therefore be omitted. Employing the identity
\begin{align}
    \int\frac{d\mathbf{n_q}}{4\pi}P_\pm(\mathbf{n_q})\sigma_\alpha P_\pm(\mathbf{n_q})=\frac{1}{4}\int\frac{d\mathbf{n_q}}{4\pi}\left(\sigma_\alpha+\sigma_\beta n_\beta\sigma_\alpha\sigma_\gamma n_\gamma\right)=\frac{1}{6}\sigma_\alpha
\end{align}
and extending the integration in Eq. \eqref{eq:ladder_calculation} to negative half axis of $\xi_\mathbf{q}$ in a way it was done in preceeding section, for one ladder step contribution we arrive at  
\begin{align}
    2\cdot\frac{1}{6}\sigma_\alpha\frac{\left(\theta(\varepsilon)-\theta(\varepsilon-\omega)\right)i/\tau^{\prime}_\eta}{i\omega+i/\tau^{\prime}_\eta}=\frac{1}{3}\sigma_\alpha\left(\theta(\varepsilon)-\theta(\varepsilon-\omega)\right)\frac{1/\tau^{\prime}_\eta}{1/\tau^{\prime}_\eta+\omega}
\end{align}
so that, after summing up the series, vertex correction factor $\Gamma(\omega)$ takes the form
\begin{align}
    \Gamma(\omega)=\hat{j}_\alpha\left(1+\frac{\frac{1}{3}\frac{1/\tau^{\prime}_\eta}{1/\tau^{\prime}_\eta-i\omega}}{1-\frac{1}{3}\frac{1/\tau^{\prime}_\eta}{1/\tau^{\prime}_\eta-i\omega}}\left(\theta(\varepsilon)-\theta(\varepsilon-\omega)\right)\right).
    \label{eq:vertex_correction}
\end{align}
Accounting for Eq. \eqref{eq:vertex_correction} leads to replacing of scattering time $\tau^{\prime}_\eta$ in Eq. \eqref{eq:chi_result} by transport time $\tau_\eta$.

\subsection{Dirac case}
Let us also note that the above calculation could be easily generalized for Dirac semimetals case. The reason is that Dirac node is actually comprised of two Weyl nodes with opposite chirality residing at the same momentum. Therefore, the Hamiltonian and Green function of a Dirac fermion read
\begin{align}
    \hat{H}_D=v_F\tau_3{\bm \sigma}\cdot\mathbf{k},\quad
    \hat{G}_D(\varepsilon)=\begin{pmatrix}
        (\varepsilon-v_F{\bm \sigma}\cdot\mathbf{k})^{-1} & 0\\
        0 & (\varepsilon+v_F{\bm \sigma}\cdot\mathbf{k})^{-1}
    \end{pmatrix},\quad \hat{\mathbf{j}}=v_F\begin{pmatrix}
        {\bm \sigma} & 0\\
        0 & -{\bm \sigma}
    \end{pmatrix}
    \label{eq:Dirac_matrices}
\end{align}
Since matrices in Eq. \eqref{eq:Dirac_matrices} are diagonal, the result Eq. (4) transfers to Dirac case as well.

\section{Nonlinear circular dichroism probe}

According to Eq. (5) of the main text, dieletric tensor is given by
\begin{align}
    \varepsilon_{ij}(\omega_0)=\begin{pmatrix}
        \varepsilon_\perp(\omega_0) & 4i\pi \sigma_{xy}(\omega_0)/\omega_0 & 0\\
        -4i\pi \sigma_{xy}(\omega_0)/\omega_0 & \varepsilon_\perp(\omega_0) & 0\\
        0 & 0 & \varepsilon_\parallel(\omega_0)
    \end{pmatrix},
    \label{eq:varepsilon_matrix}
\end{align}
where we assumed $\mathbf{E}\parallel 0z$. For the sake of model simplicity we have also assumed rotational symmetry in $0xy$ plane.

Nonlinear circular dichroism signal $\gamma=\mathrm{Im}(n_L-n_R)$ is given by
\begin{align}
    \gamma=\mathrm{Im}\left\{\sqrt{\varepsilon_\perp(\omega_0)+\frac{4\pi i\sigma_{xy}}{\omega_0}}-\sqrt{\varepsilon_\perp(\omega_0)-\frac{4\pi i\sigma_{xy}}{\omega_0}}\right\},
    \label{eq:gamma_general}
\end{align}
where 
$$
\sigma_{xy}=\frac{e^3}{24\pi^2\hbar^2}E_z\sum_\eta\frac{\eta\tau_\eta}{1-i\omega_0\tau_\eta}
$$
describes non-linear Hall conductivity. Assuming that $\sigma_{xy}\ll\omega_0$ (which in case of TaAs corresponds to THz frequencies and gives the same estimate as $\omega_0>\omega_p$ condition for non-decaying EM wave) and performing Taylor expansion of \eqref{eq:gamma_general}, we arrive at
\begin{align}
    \gamma=\frac{e^3}{6\pi\hbar^2\omega_0}E_z\sum_\eta\frac{\eta\tau_\eta}{1+\omega_0^2\tau_\eta^2}\mathrm{Re}\frac{1}{\sqrt{\varepsilon_\perp(\omega_0)}}
\end{align}


\end{widetext}

\end{document}